\documentclass[a4paper,12pt]{article}
\usepackage[left=2.5cm,bottom=3cm,right=2.5cm,top=3cm]{geometry} 

\usepackage{overpic,youngtab}
\usepackage{subfigure}
\usepackage[latin,english]{babel}
\usepackage{amsmath}
\usepackage{amssymb}
\usepackage{bigints}
\usepackage{epstopdf}
\usepackage{graphics,psfrag,rotating}
\usepackage{mathabx}
\usepackage{graphicx}
\usepackage{dcolumn}
\usepackage{float}
\usepackage{pdflscape}
\usepackage{array}
\usepackage{booktabs}
\usepackage{amscd} 
\usepackage{mathtools}
\usepackage{fancybox}
\usepackage{fix-cm}
\usepackage[colorlinks=true,citecolor=blue,,linktocpage=true,linkcolor=blue,urlcolor=black]{hyperref}
\usepackage{tikz} 
\usetikzlibrary{matrix} 
\usetikzlibrary{positioning} 
\usetikzlibrary{arrows}
\usepackage{titlesec}
\usepackage{abstract}

\def\be{\begin{equation}}
\def\ee{\end{equation}}
\def\bea{\begin{eqnarray}}
\def\eea{\end{eqnarray}}
\def\pd{\partial}

\def\a{\alpha}
\def\b{\beta}

\def\d{\delta}

\def\m{\mu}
\def\n{\nu}
\def\t{\tau}

\def\l{\lambda}

\def\r{\rho}

\def\s{\sigma}

\def\bi{\begin{itemize}}
	\def\ei{\end{itemize}}
\def\bg{\bar{g}}

\begin{document}
	
	%%%%%%%%%%%%%%%%%%%%%%%%%%%%%%%%%%%%%%%%%%%%%%%%%%%%%%%%%%%%%%%%%%%%%%%%%%%%%
	\vspace*{-1cm}
\phantom{hep-ph/***} 
{\flushleft
	{{FTUAM-24-xx}}
	\hfill{{ IFT-UAM/CSIC-24-18}}}
\vskip 1.5cm
\begin{center}
	{\LARGE\bfseries  Superposition of gravitational fields. }\\[3mm]
	\vskip .3cm
	
\end{center}

\vskip 0.5  cm
\begin{center}
	{\large Enrique \'Alvarez and  Jes\'us Anero.}
	\\
	\vskip .7cm
	{
		Departamento de F\'isica Te\'orica and Instituto de F\'{\i}sica Te\'orica, 
		IFT-UAM/CSIC,\\
		Universidad Aut\'onoma de Madrid, Cantoblanco, 28049, Madrid, Spain\\
		\vskip .1cm

		\vskip .5cm
		
		\begin{minipage}[l]{.9\textwidth}
			\begin{center} 
				\textit{E-mail:} 
				\tt{enrique.alvarez@uam.es},
				\tt{jesusanero@gmail.com} 
				%\tt{irenesanchezl2@gmail.com}
			\end{center}
		\end{minipage}
	}
\end{center}
\thispagestyle{empty}

\begin{abstract}
	\noindent
Linear superposition of gravitational fields is shown to be possible for a large class of spacetimes, in some specific gauge (frame). Explicit examples are presented.

\end{abstract}

\newpage
\tableofcontents
\thispagestyle{empty}
\flushbottom

\newpage
%%%%%%%%%%%%%%%%%%%%%%%%%%%%%%%%%%%%%%%%%%%%%%%%%%%%%%%%%%%%%%%%%%%%%%%%
\section{Introduction.}
%%%%%%%%%%%%%%%%%%%%%%%%%%%%%%%%%%%%%%%%%%%%%%%%%%%%%%%%%%%%%%%%%%%%%%%%
It is well-known that gravitational fields in the framework of General Relativity (GR), do not obey the superposition law; that is, the addition of two free gravitational fields characterized locally by two metrics, $g^{(1)}_{\m\n}(x)$  and $g^{(2)}_{\m\n}(x)$, say $g^{(12)}_{\m\n}(x)=g^{(1)}_{\m\n}(x)+g^{(2)}_{\m\n}(x)$ is not itself a free gravitational field. In the presence of an energy-momentum acting as a  source the same thing happens; the gravitational field induced by the addition of two different energy-momentum tensors $T^{(1)}_{\m\n}(x)$  and $T^{(2)}_{\m\n}(x)$, namely $T^{(12)}_{\m\n}(x)\equiv T^{(1)}_{\m\n}(x)+ T^{(2)}_{\m\n}(x) $ is not the sum of the two fields (spacetime metrics) induced  by each energy-momentum individually.
\be
g_{\m\n}\left[ T^{(1)}_{\a\b}(x)+ T^{(2)}_{\a\b}(x)\right]\neq g_{\m\n}\left[T^{(1)}_{\a\b}(x)\right]+g_{\m\n}\left[T^{(2)}_{\a\b}(x)\right]
\ee
\par
All this  of course stems from the fact that GR is a non-linear theory,  and so are Einstein's   equations of motion (EM), which reduce to Ricci flatness in the free case. The addition of two Ricci flat metrics is not itself Ricci flat in general.

\par
 One line of research \cite{Levin}\cite{Overstreet} attempted to define the combined metric $g_{\m\n}^{(12)}$ 
through some operation more complex that just a sum. It would be nice to be able to define a {\em star} operation $g^{(1)}_{\m\n}(x) \star g^{(2)}_{\m\n}(x)$ such that
\be
R_{\m\n}\left[g^{(1)}_{\a\b}(x)\star g^{(2)}_{\a\b}(x)\right]=R_{\m\n}\left[g^{(1)}_{\a\b}(x)\right]+R_{\m\n}\left[g^{(2)}_{\a\b}(x)\right]
\ee
This has however  been met with only limited success.
\par
There are exceptions however, and this  {\em in certain coordinates} only. We shall see in  detail  that in some important cases, superposition holds. Namely, there  is a class of Ricci flat spacetimes introduced by  Kerr-Schild (KS), for which G\"urses and G\"ursey \cite{Gurses} have shown that the equations of motion linearize. This class is rather large; in it one can find  some type D spacetimes, such as Kerr-Newman (and of course some important particular cases, such as Kerr, Schwarzschild and Reissner-Nordstrom)  as well as some type N spacetimes, such as gravitational plane waves.  It does not include Friedmann-Robertson-Walker spacetimes, however, which are Petrov type 0. It follows that for the KS spacetimes and in the right gauge, superposition holds.
\par
It may come  at first sight as a surprise that this superposition property only happens in certain reference frames \footnote{We shall use the words gauge, frame and coordinate choice as equivalent for the purposes of this paper.
}. If one thinks about it, it is only natural. After all, by the equivalence principle, an accelerated frame is locally equivalent to a gravitational field, so that there is always a (free falling) frame in which the gravitational field vanishes at a given point. Of course this free falling frame depends on the particular gravitational field under consideration and would be in general different for each field to be superposed.
The existence of {\em linear gauges } in our case follows the same philosophy valid in this case for all KS spacetimes.

 On the other hand, the idea that not all gauges are equivalent from the physical point of view is quite old, and can be traced back at least to Fock's work \cite{Fock}, who used to claim that physics laws hold simpler  in harmonic coordinates. That said,  KS coordinates  are not harmonic \cite{Bicak} ones in general.
\par
It should be stressed that all our results are valid in GR; no extra symmetry (e.g. supersymmetry) needs to be invoked.

%%%%%%%%%%%%%%%%%%%%%%%%%%%%%%%%%%%%%%%%%%%%%%%%%%%%%%%%%%%%%%%%%%%%%%%%
\section{When is GR linear?}
%%%%%%%%%%%%%%%%%%%%%%%%%%%%%%%%%%%%%%%%%%%%%%%%%%%%%%%%%%%%%%%%%%%%%%%%
There is an important class of metrics, namely the Kerr-Schild class \cite{Kerr}\cite{Adamo}\cite{Debney} characterized by the fact that
\be\label{basic}
g_{\m\n}=\eta_{\m\n}-2 V(x) \l_\m \l_\n
\ee
where $\eta_{\m\n} $ Minkowski's metric and the vector $\l$ is null
\be
g_{\m\n} \l^\m \l^\n=\eta_{\m\n} \l^\m \l^\n=0
\ee
This class contains many physically important type D metrics, like Kerr (and of course Schwarzschild) as well as type N plane gravitational waves. An important spacetime which does not belong to this family is Friedmann-Robertson-Walker's. The electromagnetic generalizations like Kerr-Newman are also in this class.
\par
In cartesian coordinates the metric is also unimodular
\be
|g|=1
\ee
and in fact
\be
g^{\m\n}= \eta^{\m\n}+ 2 V \l^\m \l^\n
\ee

If the vector is also  geodesic, that is
\be
\l^\m \pd_\m \l_\a=\l^\m\nabla_\m \l_\a=0
\ee
then the spacetime is algebraically special.
\par
 A straightforward calculation \cite{Gurses} shows that Ricci's tensor in the KS gauge reduces to

\bea
& R_{\m\n}=-\partial_\m\partial_\a (l^\a l_\n)-\partial_\n\partial_\a (l^\a l_\m)+\Box (l_\m l_\n)+\nonumber\\
&+2\left[\partial_\b l^\a\partial^\b l_\a-L^2+A^2-\partial_\a(K\l^\a)\right]l_\m l_\n
\eea
where
\bea
& K\equiv -\partial_\a(V\l^\a)\nonumber\\
&l_\m\equiv \sqrt{V} \l_\m\nonumber\\
&A\equiv -\l^\a \pd_\a \left(\sqrt{V}\right)\nonumber\\
&L\equiv \pd_\a l^\a
\eea
it easily follows that
\bea \label{E} &G^\m_{~\n}=\partial^\m(K\l_\n)+\partial_\n(K\l^\m)+\Box(l^\m l_\n)-\d^\m_\n\partial_\a(K\l^\a)\eea

On the other hand, the Bianchi identity implies
\be \nabla_\m G^\m_{~\b}=\partial_\m G^\m_{~\b}+\Gamma^\m_{\m\l}G^\l_{~\b}-\Gamma^\l_{\m\b}G^\m_{~\l}=0\ee
but  given that $\Gamma^\m_{\m\l}=0$  because $|g|=1$ and 
\be \Gamma^\l_{\m\b}G^\m_{~\l}=0\ee
Einstein's tensor is also conserved in the ordinary sense
\be \nabla_\m G^\m_{~\b}=\partial_\m G^\m_{~\b}=0\ee

Finally, the equation of motion reduces to
\be 2\eta^{\m\r}G^{\n}_{~\r}=\partial_\a\partial_\b\left[\eta^{\m\n}g^{\a\b}+\eta^{\a\b}g^{\m\n}-\eta^{\m\a}g^{\n\b}-\eta^{\n\a}g^{\m\b}\right]=2\kappa^2 T^{\m\n}\ee
then the Einstein's equations in the KS gauge are linear.

The Fierz-Pauli \cite{Fierz} equation (which are equivalent to the linearized Einstein's)  reads
\be
\Box \left(h_{\m\n}-h \eta_{\m\n}\right)-\pd_\m\pd^\l h_{\l\n}-\pd_\n\pd^\l h_{\l\m}+\pd_\m\pd_\n h+\eta_{\m\n}\pd^\r\pd^\s h_{\r\s}=0
\ee
(where $h\equiv \eta^{\a\b} h_{\a\b}$. The KS equation of motion is then equivalent to Fierz-Pauli's with  $h=0$ (which stems from unimodularity).
\par
In conclusion, all this means that in all KS spacetimes, there is superposition of gravitational fields, {\em provided we are using coordinates in which  the basic equation \eqref{basic} holds true}.
\par

				It is important that the reciprocal is in some sense also true (cf. the second reference in \cite{Gurses}). Given a Ricci flat spacetime with metric $\bg_{\a\b}$, and a solution of the Fierz-Pauli equations of the form
\be
h_{\m\n}=l_\m l_\n
\ee
where $l^2=0$, then 
\be
g_{\a\b}=\bg_{\a\b}+l_\a l_\b
\ee
is also Ricci flat.		

%%%%%%%%%%%%%%%%%%%%%%%%%%%%%%%%%%%%%%%%%%%%%%%%%%%%%%%%%%%%%%%%%%%%%%%%
\section{Superpositions of some vacuum spacetimes.}
%%%%%%%%%%%%%%%%%%%%%%%%%%%%%%%%%%%%%%%%%%%%%%%%%%%%%%%%%%%%%%%%%%%%%%%%
\bi
\item
 When we superpose two KS metrics we know that the sum of the two metrics will be solution of the linear equation, but {\em this linear equation will represent the Ricci tensor only when the composite metric is itself KS}, which will usually not be the case.
 Let us study this topic in a bit more detail. Assume two KS metrics

\bea & \text{d}s^2_{(V)}=\eta_{\m\n}\text{d}x^\m\text{d}x^\n-2V(x)\l_\m \l_\n\text{d}x^\m\text{d}x^\n\nonumber\\
& \text{d}s^2_{(F)}=\eta_{\m\n}\text{d}x^\m\text{d}x^\n-2F(x)\t_\m \t_\n\text{d}x^\m\text{d}x^\n
\eea
where $\eta_{\m\n}$ is the Minkowski metric ,  $V$ and $F$ are  scalar functions and  $\l_\m$ and $\t_\m$ are null vectors.
The sum of the metrics reads
\bea  \text{d}s^2_{(V+F)}&=\eta_{\m\n}\text{d}x^\m\text{d}x^\n-\left(V\l_\m \l_\n+F\t_\m \t_\n\right)\text{d}x^\m\text{d}x^\n\eea
after a rescaling in order to fix the asymptotic behavior
\be x^\m\rightarrow \frac{1}{\sqrt{2}}x^\m\ee
of course, $V$, $F$, $\l_\m$ and $\t_\m$ are rescaled too, but this detail is not interesting for our demostration

There are several conditions to be met in order for the sum of two KS metrics to be also KS

To begin with, note the inverse metric reads
\be g^{\m\n}=\eta^{\m\n}-\left(V\l^\m \l^\n+F\t^\m \t^\n\right)\ee
only if
\be\l_\m \t^\m=0\ee

On the other hand, we need to  assume that both vector fields are geodesic
\bea &\l^\m\partial_\m \t_\n=0\nonumber\\
&\t^\m\partial_\m \l_\n=0
\eea
Then the connection reads

\bea &\Gamma^\a_{\m\n}(V+F)=\frac{1}{2}\Gamma^\a_{\m\n}(V)+\frac{1}{2}\Gamma^\a_{\m\n}(F)+\frac{1}{2}A^\a_{\m\n}\eea
with
\be A^\a_{\m\n}\equiv V\l^\a\l^\r\t_\m\t_\n\partial_\r F +F\t^\a\t^\r\l_\m\l_\n\partial_\r V-V\l^\a\l^\r\l_\m\l_\n\partial_\r V -F\t^\a\t^\r\t_\m\t_\n\partial_\r F\ee
%in consequence, again with $\l_\m \t^\m=0$
%\bea &\Gamma^\m_{\m\n}=\frac{1}{2}\Gamma^\m_{\m\n}(V)+\frac{1}{2}\Gamma^\m_{\m\n}(F)=0\eea

and the Ricci tensor
\bea
&&R_{\m\n}^{(V+F)}=\frac{1}{2}R_{\m\n}^{(V)}+\frac{1}{2}R_{\m\n}^{(F)}+\frac{1}{2}\partial_\r A^\r_{\m\n}+\nonumber\\
&&+\frac{1}{4}\left[\Gamma^\l_{\m\r}(V)\Gamma^\r_{\l\n}(V)+\Gamma^\l_{\m\r}(F)\Gamma^\r_{\l\n}(F)-\Gamma^\l_{\m\r}(V)\Gamma^\r_{\l\n}(F)-\Gamma^\l_{\m\r}(F)\Gamma^\r_{\l\n}(V)\right]
\eea
%where in general
%\bea \partial_\r A^\r_{\m\n}&=\l^\a\t_\m\t_\n\partial_\r(V\l^\r\partial_\a F)+\t^\a\l_\m\l_\n\partial_\r(F\t^\r\partial_\a V)-\nonumber\\
%&-\l^\a\l_\m\l_\n\partial_\r(V\l^\r\partial_\a V)-\t^\a\t_\m\t_\n\partial_\r(F\t^\r\partial_\a F)\eea
To conclude, in general, \textit{the Ricci tensor of the sum of metrics is not the sum of the individual Ricci tensors}. A sufficient condition for additivity  is  $V=F$ and $\l_\m=\t_\m$ (then $A^\r_{\m\n}=0$).

%%%%%%%%%%%%%%%%%%%%%%%%%%%%%%%%%%%%%%%%%%%%%%%%%%%%%%%%%%%%%%%%%%%%%%
\item
In order to stress the importance of the reference system, let us  consider the simplest example of Schwarzschild space time  in the KS gauge where

\be
V={r_s \over 2 r }
\ee
and 
\be
\l_\m dx^\m= dt +{x dx + y dy+ z dz\over r}= dt+dr
\ee
We use, $r=\sqrt{x^2+y^2+z^2}$, only as an abbreviation, not as a coordinate. To be specific, the form of the metric in the KS gauge reads
%The Schwarzschild in Kerr-Schild coordinates
%\be \text{d}s^2=\text{d}t^2-\text{d}r^2-r^2\text{d}\Omega^2-\frac{r_s}{r}\left(\text{d}t^2+\text{d}r^2+2\text{d}r\text{d}t\right)\ee
%with $l_\m=(1,1,0,0)$ and $l_\m l^\m= 0$, this metric is Ricci flat.

%The superposition, with different Schwarzschild radii
%	\bea \text{d}s^2_{(1)}&=\text{d}t^2-\text{d}r^2-r^2\text{d}\Omega^2-\frac{r_1}{r}\left(\text{d}t^2+\text{d}r^2+2\text{d}r\text{d}t\right)\nonumber\\
%	\text{d}s^2_{(2)}&=\text{d}t^2-\text{d}r^2-(r-r_0)^2\text{d}\hat{\Omega}^2-\frac{r_2}{r-r_0}\left(\text{d}t^2+\text{d}r^2+2\text{d}r\text{d}t\right)\eea
%where
%\be r-r_0\equiv \sqrt{(x-x_0)^2+(y-y_0)^2+(z-z_0)^2}\ee
%these metrics have different light vector $l_\m=(1,1,0,0)$ and $t_\m=(1,1,f(r,\theta,\varphi),g(r,\theta,\varphi))$  but $\eta^{\m\n}l_\m t_\n=0$, both metrics are Ricci flat,  finally the superposition
%\bea \text{d}s^2_{(1+2)}&=\text{d}t^2-\text{d}r^2-r^2\text{d}\Omega^2-\frac{r_1}{r}\left(\text{d}t^2+\text{d}r^2+2\text{d}r\text{d}t\right)+\nonumber\\
%&+\text{d}t^2-\text{d}r^2-(r-r_0)^2\text{d}\hat{\Omega}^2-\frac{r_2}{r-r_0}\left(\text{d}t^2+\text{d}r^2+2\text{d}r\text{d}t\right)\eea
%is Ricci flat.

%Different case is in Cartesian coordinates
%\bea \text{d}s^2_{(1)}&=\left(1-\frac{r^{(1)}_s}{r}\right)\text{d}t^2-\left(\d_{ij}+\frac{r^{(1)}_s}{r^3}\sum_{ij}x^ix^j\right)\text{d}x^i\text{d}x^j-\frac{2r^{(1)}_s}{r^2}\sum_{i}x^i\text{d}x^i\text{d}t=\nonumber\\
%&=\eta_{\m\n}dx^\m dx^n-\frac{r^{(1)}_s}{r^3}\l_\m\l_\n dx^\m dx^n\eea
\bea \label{KS} &&\text{d}s^2_{(1)}=\eta_{\m\n}dx^\m dx^n-\frac{r^{(1)}_s}{r^3}\l_\m\l_\n dx^\m dx^n\eea
this metric has a light vector
%\be \l_\m=\left(1,\frac{x}{r},\frac{y}{r},\frac{z}{r}\right)\ee
\be \l_\m=\left(r,x,y,z\right)\ee
It can be easily verified that this metric is Ricci flat and unimodular.

%is Ricci flat.
%\be\label{KS}
%ds^2=\left(1-{r_s\over r}\right)dt^2-\left(\d_{ij}+{r_s\over r^3} \sum_{ij}x^i x^j \right)dx^i dx^j-2 r_s\sum_i {x^i\over r^2} dx^i dt
%\ee

\par
Let us now consider a superposition of two such metrics
\bea \text{d}s^2_{(1+2)}&=\eta_{\m\n}dx^\m dx^n-\frac{r^{(1)}_s+r^{(2)}_s}{\sqrt{2}r^3}\l_\m\l_\n dx^\m dx^n\eea
It is possible to check that this superposition is Ricci flat as well.
\par

We learn that the Schwarzschild radius of the superposition is
\be
r_s^{(12)}={r_s^{(1)}+r_s^{(2)}\over \sqrt{2}}
\ee
This of course, the same relationship of the mass of the combined spacetime containing two black holes on top of each other with the parent masses. It is to be remarked that this mass is {\em smaller} than the sum of the two initial masses.
 The entropy, in turn, is
 \be
 S_{(12)}=\pi \left(r_s^{ (12)}\right)^2=\pi {\left(r_s^{(1)}+r_s^{(2)}\right)^2\over 2}=S_{(1)}+S_{(2)} -{\pi\over 2} \left(r_s^{(1)}-r_s^{(2)}\right)^2
\ee
which again, {\em smaller} than the sum of the two parent entropies. We do not know any physical interpretation of this fact. Were we to add $n$ copies of the BH, a similar rescaling leas to 
\be
r_s^{(n)}={1\over \sqrt{n}}\sum_{i=1}^n r_s^{(i)}
\ee
again,  {\em smaller} than the sum of the two initial masses.
 The entropy reads
 \be
 S_{(n)}=\pi \left(r_s^{ (n)}\right)^2=\sum_{i=1}^nS_{(i)}-\pi {n-1\over n}\sum_{i=1}^n (r_s^{(i)})^2+{\pi\over n} \prod_{i\neq j} r_s^{(i)}r_s^{(j)}
\ee

\item Let us consider  the case when one of the black holes is displaced, so the their  singularites do not coincide.

	Assume a displaced metric centered  at a  point $\vec{r}=\vec{a}=(a_x,a_y,a_z)$
	%\bea &\text{d}s^2_{(2)}=\eta_{\m\n}dx^\m dx^n-\frac{r^{(1)}_s}{\sqrt{r^2-2\vec{a}\cdot\vec{r}+\vec{a}^2}}\t_\m\t_\n dx^\m dx^n\eea
	\bea &\text{d}s^2_{(2)}=\eta_{\m\n}dx^\m dx^n-\frac{r^{(2)}_s}{\left(r^2-2\vec{a}\cdot\vec{r}+\vec{a}^2\right)^{3/2}}\t_\m\t_\n dx^\m dx^n\eea
	%\bea &\text{d}s^2_{(2)}=\left(1-\frac{r^{(2)}_s}{\sqrt{r^2-2\vec{a}\cdot\vec{r}+\vec{a}^2}}\right)\text{d}t^2-\left(\d_{ij}+\frac{r^{(2)}_s}{\left(r^2-2\vec{a}\cdot\vec{r}+\vec{a}^2\right)^{3/2}}\sum_{ij}x^ix^j\right)\text{d}x^i\text{d}x^j-\nonumber\\
	%&-\frac{2r^{(2)}_s}{\left(r^2-2\vec{a}\cdot\vec{r}+\vec{a}^2\right)}\sum_{i}x^i\text{d}x^i\text{d}t\eea
	This metric has a light vector
	%\be t_\m=\left(1,\frac{x-a_x}{\sqrt{r^2-2\vec{a}\cdot\vec{r}+\vec{a}^2}},\frac{y-a_y}{\sqrt{r^2-2\vec{a}\cdot\vec{r}+\vec{a}^2}},\frac{z-a_z}{\sqrt{r^2-2\vec{a}\cdot\vec{r}+\vec{a}^2}}\right)\ee
	\be \t_\m=\left(\sqrt{r^2-2\vec{a}\cdot\vec{r}+\vec{a}^2},x-a_x,y-a_y,z-a_z\right)\ee	
Now the superposition of both metrics centered at the origin and the displaced one (with the usual rescaling)
\bea \text{d}s^2_{(1+2)}&=\eta_{\m\n}dx^\m dx^n-\frac{r^{(1)}_s}{\sqrt{2}r^3}\l_\m\l_\n dx^\m dx^n-\frac{r^{(2)}_s}{\sqrt{2}\left(r^2-2\vec{a}\cdot\vec{r}+\vec{a}^2\right)^{3/2}}\t_\m\t_\n dx^\m dx^n\eea but in this metric $\eta^{\m\n}\l_\m \t_\n\neq 0$, and the superposition is not possible.

 \item In the usual  Schwarzschild coordinates
\be
\text{d}s^2_{(1)}=\left(1-\frac{r_s^{(1)}}{r}\right)\text{d}t-\frac{1}{\left(1-\frac{r_s^{(1)}}{r}\right)}\text{d}r-\text{d}\Omega^2
\ee
and
\be
\text{d}s^2_{(2)}=\left(1-\frac{r_s^{(2)}}{r}\right)\text{d}t-\frac{1}{\left(1-\frac{r_s^{(2)}}{r}\right)}\text{d}r-\text{d}\Omega^2
\ee
%for simplicity we call $r_s^{(1)}$ and $r_s^{(2)}$ the Schwarzschild radii. 
\par
In spite of the fact that both metrics are Ricci flat,  their sum
\be
\text{d}s^2_{(12)}=\left(2-\frac{r_s^{(1)}+r_s^{(2)}}{r}\right)\text{d}t-\left[\frac{1}{\left(1-\frac{r_s^{(1)}}{r}\right)}+\frac{1}{\left(1-\frac{r_s^{(2)}}{r}\right)}\right]\text{d}r-2\text{d}\Omega^2
\ee
is not.
This means that there is no superposition in this gauge.

\item In fact  one can transform   the usual reference gauge to the  Kerr-Schild one as follows. Start from
\be \text{d}s^2=\left(1-\frac{r_s}{r}\right)\text{d}t^2-\left(1+\frac{r_s}{r}\right)\text{d}r^2-\frac{2r_s}{r}\text{d}r\text{d}t-r^2\text{d}\Omega^2\ee
Define a new temporal coordinate
\be\label{w} \text{d}\bar{t}=f\left[A\text{d}t-B\text{d}r\right]\ee
Then
\be A\text{d}t^2-2B\text{d}t\text{d}r=\frac{\text{d}\bar{t}^2}{f^2A}-\frac{B^2\text{d}r^2}{A}\ee
where in our case
\bea
&&A=1-\frac{r_s}{r}\nonumber\\
&&B=\frac{r_s}{r}
\eea
and the metric reduces to
\be \text{d}s^2=\frac{\text{d}\bar{t}^2}{f^2\left(1-\frac{r_s}{r}\right)}-\frac{1}{\left(1-\frac{r_s}{r}\right)}\text{d}r^2-r^2\text{d}\Omega^2\ee
Taking $f^2\equiv \frac{1}{\left(1-\frac{r_s}{r}\right)^2}$ (this makes the second member of  \eqref{w} a total differential) yields  the traditional metric in Schwarzschild coordinates
\be \text{d}s^2=\left(1-\frac{r_s}{r}\right)\text{d}\bar{t}^2-\frac{1}{\left(1-\frac{r_s}{r}\right)}\text{d}r^2-r^2\text{d}\Omega^2\ee
\item 
In fact, it is even simpler in Eddington-Filkenstein outgoing coordinates, where
\be
u=t-r^*
\ee
and 
\be
r^*\equiv r+ r_s \log\,(r-r_s)
\ee
 is the usual tortoise coordinate, Schwarzschild's metric reads
 \be\label{ef}
\text{d}s^2= \text{d}u^2+ 2 \text{d}u \text{d}r -r^2 \text{d}\Omega^2-{r_s\over r} \text{d}u^2
 \ee
 which is already in the KS gauge with
 \bea
 &&2 V={r_s\over r}\nonumber\\
 &&\l_\m\text{d} x^\m=\text{d} u
 \eea
 
  It is however plain that this form is not unimodular. This can be fixed by defining new coordinates $\m$ and $x$ such that
 \bea
 &x={r^3\over 3}\nonumber\\
 &\m=\cos\,\theta
 \eea
 The unimodular metric reads
 \be\label{uni}
 ds^2= \left(1-{r_s\over (3x)^{1/3}}\right)du^2-2 (3 x)^{-2/ 3} du dx -(3x)^{2/3}\left({d\m^2\over 1-\m^2}+(1-\m^2) d\phi^2\right)
 \ee
 It is remarkable that superposition works with both metrics  \eqref{ef} and \eqref{uni}.

\item Lest one thinks that superposition only happens when the metric is already linear in the external parameters, let us consider the Kerr-Newman spacetime whose metric reads \cite{Adamo}
\be
g_{\m\n}= \eta_{\m\n}- 2 V \l_\m\l_\n
\ee
with 
\be
2 V={2 G m r^3-Q^2 r^2\over r^4 + a^2 z^2}
\ee
and 
\be
\l_\m dx^\m= dt +{z\over r} dz+{r\over r^2+a^2}(x dx + y dy)-{a\over r^2+a^2}(x dt-y dx)
\ee
where $r$ is defined through
\be
{x^2+y^2\over r^2+a^2}+{z^2\over r^2}=1
\ee
Here the parameter $m$ defines the mass of the spacetime, $a$ is related to the angular momentum, and $Q$ stands for  the electromagnetic charge. The mass is the only parameter which appears in a linear way.
\par
When superposing $n$ such spacetimes we find after rescaling

\bea
&&r^2_{(n)}={r^2\over n}\nonumber\\
&&a^2_{(n)}={a^2\over n}
\eea
This means that

\be
2 V^{(n)}={2\sqrt{n} G m r^3-n Q^2 r^2\over r^4 + a^2 z^2}
\ee
and 

\be
\l^{(n)}_\m dx^\m=\frac{1}{\sqrt{n} }\l_\m dx^\m
\ee
\par
The superposition of $n$ such Kerr-Newman spacetimes is another Kerr-Newman spacetime, with the parameters changed as shown, and this in spite of the explicit non-linearity of the dependence in said parameters. That is
\be
g^{(n)}_{\m\n}=\eta_{\m\n}- 2 V^{(n)}\l^{(n)}_\m\l^{(n)}_\n
\ee
%%%%%%%%%%%%%%%%%%%%%%%%%%%%%%%%%%%%%%%%%%%%%%%%%%%%%%%%%%%%%%%%%%%%%
\item Consider now gravitational plane waves 
\be
\text{d}s^2=\text{d}u\text{d}v-H_{ab}[u]x^ax^b\text{d}u^2-\d_{ab}\text{d}x^a\text{d}x^b
\ee
which  are Ricci flat if $H_{ab}$ is traceless (cf. \cite{Alvarez} for a recent reference)
\be H_{pq}[u]x^px^q\equiv a_{+}[u](x^2-y^2)+2a_{\times}[u]xy\ee
\par
They are already in the KS gauge with
\be
\l_\m dx^\m= du
\ee
It is plain that the sum of $n$ such Ricci flat waves is also another Ricci flat  wave of the same type. After rescaling
\be
\text{d}s_{(n)}^2=\text{d}u\text{d}v-\sum_{i=1}^n {H^{(i)}_{ab}[u/\sqrt{n}]\over n^2}\,x^ax^b\text{d}u^2-\d_{ab}\text{d}x^a\text{d}x^b
\ee

In fact the same is true for  the more general pp waves
\be
\text{d}s^2=\text{d}u\text{d}v-F[u,x^c]\text{d}u^2-\d_{ab} \text{d}x^a\text{d}x^b
\ee
which share the same null vector $\l_\m$.

\ei
%%%%%%%%%%%%%%%%%%%%%%%%%%%%%%%%%%%%%%%%%%%%%%%%%%%%%%%%%%%%%%%%%%%%%%%
\section{Conclusions}
%%%%%%%%%%%%%%%%%%%%%%%%%%%%%%%%%%%%%%%%%%%%%%%%%%%%%%%%%%%%%%%%%%%%%
In spite of the fact that GR is a non-linear theory,  and so are Einstein's   equations of motion (EM), which reduce to Ricci flatness in the free case, in this paper we have pointed out that in some cases one can superpose gravitational fields, but  this  {\em in certain coordinates} only. This happens in a class of  spacetimes introduced by  Kerr-Schild, for which  Einstein's equations  linearize. This class is rather large; in it one can find  some type D spacetimes, such as Kerr-Newman (and of course some important particular cases, such as  Kerr, Schwarzschild and Reissner-Nordstrom)  as well as some type N spacetimes, such as gravitational plane waves.  It follows that for these spacetimes and in the right coordinates, superposition holds.
\par
We have also discussed why it is not physically surprising that  this superposition property only takes place  in certain reference frames. In fact, it is only natural. Not all frames are equivalent from the physical point of view. After all, by the equivalence principle, an accelerated frame is locally equivalent to a gravitational field, so that there is always a (free falling) frame in which the gravitational field vanishes at a given point. Of course this free falling frame depends on the particular gravitational field under consideration and would be in general different for each field to be superposed. 
\par
The existence of {\em linear gauges} in our case follows the same philosophy valid in this case for all Kerr-Schild spacetimes; this is the main surprising result. In some sense, the KS gauge is the generalization of the free falling frame valid for all spacetimes in this family. In this gauge and for those spacetimes, gravity is linear. We then learn that superposition in this framework is not a diffeomorphism invariant property. 
\par
The superposition property we have studied in this paper seems exceedingly useful from the physical point of view as a means to get new solutions out of old ones, provided some conditions are met, which we have studied in detail. This procedure is much simpler in the free case because then we do not have to worry about sources; all solutions are simply Ricci flat. Needless to say, once new solutions are obtained, they can be transformed to any convenient new gauge.

\par
We are still a long way from being able to elucidate the physical implications of our results, however. One of the most important questions is what is the meaning of the KS gauge in the general case and how it can it be related to physical  observables. 
Some quantum effects in these spacetimes have been studied in \cite{Alvarez}. It would be very interesting to see whether some generalization of wave packets is possible for gravitational plane waves.

It has been also pointed out that all our results are valid in GR; no extra symmetry (e.g. supersymmetry) needs to be invoked.

%%%%%%%%%%%%%%%%%%%%%%%%%%%%%%%%%%%%%%%%%%%%%%%%%%%%%%%%%%%%%%%%%%%%%%%%
\section{Acknowledgements.}
%%%%%%%%%%%%%%%%%%%%%%%%%%%%%%%%%%%%%%%%%%%%%%%%%%%%%%%%%%%%%%%%%%%%%%%%

 One of us (EA) is grateful for  stimulating discussions with Andy Cohen and Yi Wang. We are grateful to Eduardo Velasco-Aja for help with the computations. We acknowledge partial financial support by the Spanish MINECO through the Centro de excelencia Severo Ochoa Program  under Grant CEX2020-001007-S  funded by MCIN/AEI/10.13039/501100011033.
 We also acknowledge partial financial support by the Spanish Research Agency (Agencia Estatal de Investigaci\'on) through the grant PID2022-137127NB-I00 funded by MCIN/AEI/10.13039/501100011033/ FEDER, UE.
All authors acknowledge the European Union's Horizon 2020 research and innovation programme under the Marie Sklodowska-Curie grant agreement No 860881-HIDDeN and also byGrant PID2019-108892RB-I00 funded by MCIN/AEI/ 10.13039/501100011033 and by ``ERDF A way of making Europe''.

\newpage
%%%%%%%%%%%%%%%%%%%%%%%%%%%%%%%%%%%%%%%%%%%%%%%%%%%%%%%%%%%%%%%%%%%%%%%%%%%%%%%%%%%%%%%%%%%%%%%%%%%%%%%%%%%%%%%%%%%%%%%%%%%%%
  
\end{document}